\documentstyle[twocolumn,epsf]{jpsj}
\newcommand{\figwidth}{8.5cm}
\newcommand{\vup}{\vspace*{-1pc}}

\def\runtitle{Random Fixed Point of Three-Dimensional Random-Bond Ising
Models }
\def\runauthor{Koji {\sc Hukushima}}

\title{
Random Fixed Point of Three-Dimensional Random-Bond Ising  Models
}
\author{
Koji {\sc Hukushima}
}
\inst{
Institute for Solid State Physics, Univ. of Tokyo, 7-22-1
  Roppongi, Minato-ku, Tokyo 106-8666
}
\recdate{November 12, 1999}

\abst{
The fixed-point structure of three-dimensional bond-disordered Ising models
 is investigated using the numerical domain-wall renormalization-group
 method.  
It is found that, in the $\pm J$ Ising model, there exists a non-trivial
 fixed point along the phase boundary between the paramagnetic and
 ferromagnetic  phases. The fixed-point  Hamiltonian of the $\pm J$ model
 numerically  coincides with  that of the unfrustrated random Ising
 models,  strongly suggesting that both belong to the same universality class. 
Another fixed point corresponding to the multicritical point is also
 found  in the $\pm J$ model.  
Critical properties associated with the fixed point are qualitatively
 consistent with theoretical predictions. 
}
\kword{domain wall renormalization group, fixed point, random system, 
spin glass, MC simulation}

\begin{document}
\sloppy
\maketitle

The influence of quenched disorder on  model systems has attracted 
considerable  interest in the field of statistical physics. 
The first remarkable criterion was given by Harris\cite{Harris}, who
claimed that if the specific-heat exponent $\alpha_{\rm pure}$ of the
pure system is positive, disorder becomes  relevant, implying that a
new random fixed point governs critical phenomena of the random
system.  One of the simplest models belonging to such a class is the
three-dimensional ($3D$) Ising model. 
Critical exponents associated with the random fixed point have been
investigated for dilution-type disorder by 
experimental\cite{Experiment1,Experiment2,Experiment3},
theoretical\cite{GrinsteinLuther,Janssen} and  numerical
approaches\cite{Heuer,BFMMRG}.  
Recently, extensive Monte Carlo (MC) studies\cite{BFMMRG}
 have clarified that the critical exponents of the $3D$ site-diluted
 Ising model are independent of the concentration of the site dilution
 $p$,  suggesting the existence of a random fixed point. 
These results were obtained by carefully taking into account correction
for finite-size scaling, unless the exponents clearly depended 
on $p$\cite{BFMMRG}. 
Experimentally, the critical exponents of randomly diluted
antiferromagnetic Ising  compounds,
Fe$_x$Zn$_{1-x}$F$_2$\cite{Experiment2,Experiment3} and 
Mn$_x$Zn$_{1-x}$F$_2$\cite{Experiment1},  are distinct from those of
the pure $3D$ Ising model. 

While the  existence of the random fixed point has been established for
the $3D$ site-diluted Ising model, the idea of the universality class
for random systems, namely classification by fixed points, has not
been explored yet as compared with various pure systems. 
In particular, the question as to whether the random fixed point is
universal irrespective of the type of disorder or not is a non-trivial
problem.  
In the present work, we study critical phenomena associated with the
ferromagnetic phase transition in $3D$ site- and bond-diluted and $\pm
J$ Ising models.  
The main purpose is to determine the fixed-point structure of $3D$
random-bond Ising models by making use of a numerical
renormalization-group (RG) analysis. 
Our strategy is based on the domain-wall  RG (DWRG) method proposed by
McMillan\cite{McMillan,McMillan2}. 
This method has been applied to a $2D$ frustrated random-bond Ising
model\cite{McMillan,McMillan2} where there is no random fixed point, and
recently 
to a $2D \pm J$ frustrated random-bond three-state Potts model\cite{SGH}
which displays  a non-trivial random fixed point.  
In this paper, we show systematic RG flow diagrams for $3D$ 
random Ising spin systems,
 which convinces us of the existence of the random fixed point.

Let us first explain briefly the RG scheme and the numerical method used 
by us. 
In the DWRG\cite{McMillan,McMillan2}, 
the following domain-wall free energy $\Delta F_J$ of a spin system on a
cube with the size $L$  is regarded as an effective coupling associated
with a length scale $L$ for a particular bond configuration, denoted by
$J$; 
\begin{equation}
\frac{\Delta F_J(T)}{T} = \ln \frac{Z_{\rm P}(T)}{Z_{\rm AP}(T)},
\end{equation}
where $Z_{\rm P(AP)}(T)$ is the partition function of the cube at
temperature $T$ under (anti-) periodic boundary conditions for a given
direction,  while the periodic boundary conditions are imposed for the
remaining directions. 
In disordered systems, the distribution $P(\Delta F)$ of the
effective couplings over the bond configurations is considered to be a
relevant quantity.  
Therefore, in the DWRG scheme, we are interested in how the distribution
is  renormalized as $L$ increases. 
In the ferromagnetic phase, the expectation value of the distribution
approaches infinity as $L$ increases, while it vanishes in the paramagnetic
phase.  
Fixed points are characterized by an invariant distribution of the
coupling under a DWRG transformation, namely, increasing $L$. 
For example, the unstable fixed-point distribution corresponding to the
pure-ferromagnetic phase transition is a delta function with a non-zero
mean. The random fixed point, if any, is expected to have a 
non-trivial distribution with a finite width.  

It is convenient to consider typical quantities characterizing the
distribution $P(\Delta F)$, instead of the distribution itself. 
We discuss here the mean $\overline{\Delta F}$ and the width
$\sigma(\Delta F)$ of the distribution which are evaluated as
\begin{eqnarray}
 \overline{\Delta F} & = & \int \mbox{d} \Delta F\  \Delta F\  P(\Delta F), \\
\sigma^2 (\Delta F) & = &  \overline{\Delta F^2} - \overline{\Delta F}^2,
\end{eqnarray}
respectively. 
Using these quantities, we define two reduced parameters, 
$r=\sigma(\Delta F)/\overline{\Delta F}$  and $t=T/\overline{\Delta F}$, 
as in the previous works\cite{McMillan,McMillan2,SGH}. 
The renormalized parameters $(r(L),t(L))$ of the length scale $L$ are
estimated numerically with bare parameters in a model Hamiltonian fixed. 
Then, the RG flow is represented by an arrow connected from the point
$(r(L),t(L))$ of size $L$ to $(r(L'),t(L'))$ of a larger size $L'$. 
Fixed points should be observed as points where the position is
invariant under the RG transformation. 
Linearizing the flow about a fixed point $(r^*,t^*)$, we obtain 
\begin{equation}
 \left(
\matrix{r(L')-r^* \cr t(L')-t^* }
\right)
=\hat{T}
\left(
\matrix{r(L)-r^* \cr t(L)-t^* }
\right),
\end{equation}
where $\hat{T}$ is a RG transformation matrix whose eigenvectors are
scaling axes of the RG flow. 
The matrix elements of $\hat{T}$ as well as the fixed point $(r^*,t^*)$
can be determined by least-squares fitting from the data point
numerically obtained. 
The critical exponents $y$ are  obtained as $\log\lambda/\log(L'/L)$ 
with the eigenvalue $\lambda$ of $\hat{T}$. 

We consider Ising models with quenched disorder which is defined
on a simple cubic lattice. The model Hamiltonian is 
\begin{equation}
 \label{eqn:model}
    {\cal H}(J_{ij},S_i)  =  -\sum_{\langle ij\rangle}J_{ij}S_iS_j, 
\end{equation} 
where the sum runs over nearest-neighbor sites. 
In order to obtain the domain-wall free energy, we use a recently
proposed MC method\cite{KH} which enables us to evaluate the
free-energy {\it difference} directly with sufficient accuracy. 
In the novel MC algorithm, 
we introduce a dynamical variable specifying the boundary conditions,
which is the sign of the interactions between the first and the last
layer of the cube for the given direction. 
The algorithm is based on the exchange MC method\cite{EMC},
sometimes called the parallel tempering\cite{Marinari}, which turns out
to be reasonably efficient for randomly frustrated spin systems. 
\begin{figure}
\epsfxsize=\figwidth
\epsffile{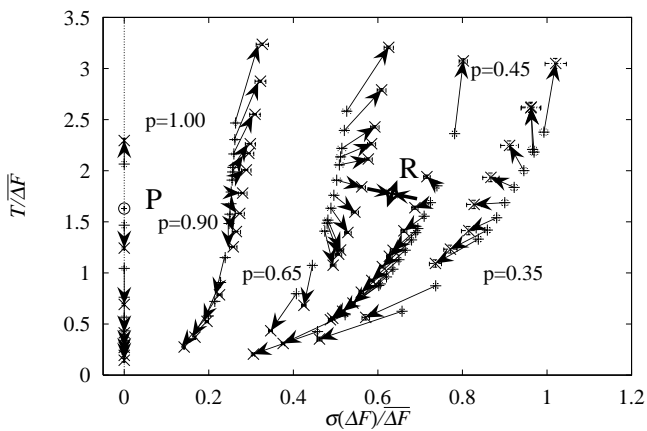}
\vup
\caption{
Flow diagram for the $3D$ unfrustrated bond-diluted Ising model. 
The random fixed point is found numerically at $(0.63(2), 1.77(5))$. 
The bold arrows indicate the eigenvectors, whose 
exponents are $y_1=1.47(4)$ and $y_2=-1.3(4)$. 
}
\vup
\label{fig:flow-bond}
\end{figure}

First we study a $3D$ bond-diluted Ising model, which is given by
eq.~(\ref{eqn:model}) with the bond distribution, $P(J_{ij}) = p\delta
(J_{ij}-J)+(1-p)\delta (J_{ij})$.   
The bond concentrations studied are $p=1.00$, $0.90$, $0.65$, $0.45$ and
$0.35$ with the system sizes $L=8$ and $L=12$. 
Sample averages are taken over $128-1984$ samples depending on the size
and the concentration. 
Errors are estimated from statistical fluctuation over samples.
The number of temperature points in the exchange MC is
fixed at $20$. 
We distribute these temperatures to replicas for each $p$ such that the
acceptance ratio for each exchange process becomes constant. 
We show the RG flow diagram in Fig. \ref{fig:flow-bond}.
When disorder is absent, corresponding to the $x=0$ axis in
Fig.~\ref{fig:flow-bond}, the pure fixed point is observed at 
$(0,1.63)$, denoted by $P$. 
Near the fixed point $P$, the arrows flow away from $P$ as disorder
is introduced. This means that the pure fixed point of the $3D$ Ising
model is unstable against the disorder, consistent with the Harris
criterion\cite{Harris}. 
Meanwhile, this finding suggests that 
a characteristic feature of the RG flow is reproduced within the system
sizes studied. 
Apart from the pure fixed point $P$, we find another fixed point,
denoted by $R$, along the phase boundary. 
The RG flows approach $R$ from both sides, indicating that it is an
attractive fixed point along the critical surface. 
The fixed point $R$ governs critical phenomena of the disordered
ferromagnetic phase transition. Therefore, it should be called the
random fixed point. 
The position of $R$ in the parameter space is obtained numerically as
$(0.63(2),1.77(5))$. At $R$, the exponents $y_{1}=1.47(4)$, $y_{2}=-1.3(4)$
and the corresponding eigenvectors are indicated by the bold arrows in
Fig.~\ref{fig:flow-bond}.
The critical exponent $\nu$, the inverse of the larger eigenvalue $y_1$, 
 is compatible with that of the site-diluted Ising model\cite{BFMMRG},
 although our estimate is not so accurate. 
All the arrows below $T_{\rm c}$ go to the unique ferromagnetic fixed
point at $(0,0)$. 
\begin{figure}
\epsfxsize=\figwidth
\epsffile{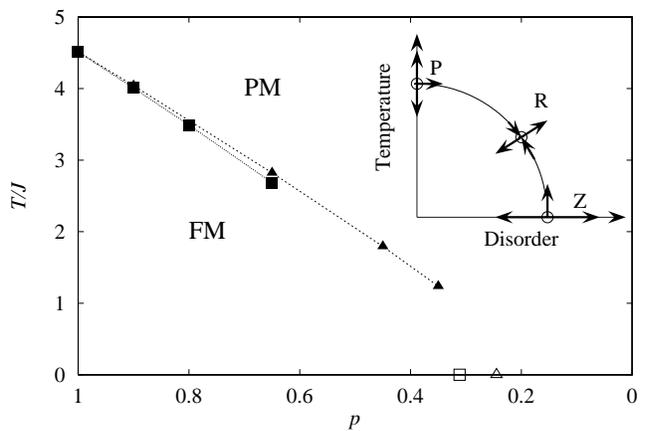}
\vup
\caption{
Phase diagram of the $3D$ random-bond Ising models 
with the bond (triangle) and the site (box) dilution.  
In the inset, a schematic flow diagram for the diluted Ising models 
is shown as a function of temperature and disorder. 
The fixed points and the RG flow are indicated.
}
\vup
\label{fig:pd-dilute}
\end{figure}

The transition temperatures are estimated by a naive finite-size scaling
assumption\cite{McMillan,McMillan2}, 
$\overline{\Delta F}=f((T-T_{\rm c})L^{1/\nu})$ with the
correlation-length exponent $\nu$,  
namely, the crossing point of $\overline{\Delta F}$ with $L=8$ and $12$
as a function of temperature is located on $T_{\rm c}$. 
This scaling form is similar to that of the Binder parameter frequently
used. 
The estimated $T_{\rm c}$ is shown in Fig.~\ref{fig:pd-dilute}. 
The exponent $\nu$ obtained by the scaling depends significantly on the
concentration $p$, similar to those observed as an 
effective exponent in the MC simulation of the site-disordered Ising
model\cite{BFMMRG}. 
It is found from the RG flow diagram that the system has a subleading
scaling parameter which gives rise to a systematic correction to the
scaling. 
Therefore, we conclude that the continuously varying exponent observed
in the naive scaling is due to the subleading parameter. 
\begin{figure}
\epsfxsize=\figwidth
\epsffile{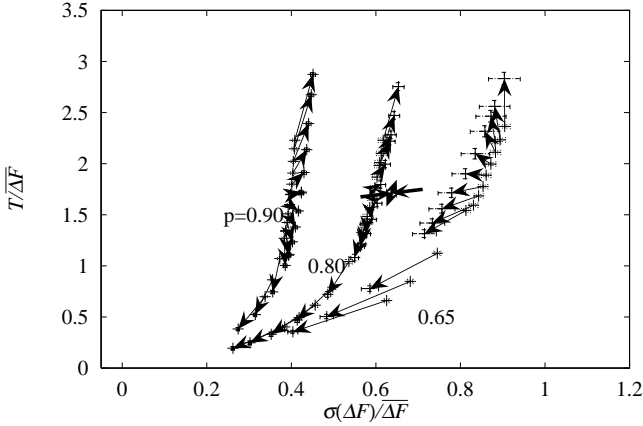}
\vup
\caption{
Flow diagram for the $3D$ unfrustrated site-diluted Ising model. 
The random fixed point is located on the point $(0.64(4), 1.71(9))$, whose
exponents are $y_1=1.37(9)$ and $y_2=-0.9(5)$. 
}
\vup
\label{fig:flow-site}
\end{figure}

We also investigate a $3D$ site-diluted Ising model by the same
procedure as in the bond-diluted Ising model described above. The model
is given by eq.~(\ref{eqn:model}) but with the bond
$J_{ij}=J\epsilon_i\epsilon_j$ 
and the distribution, $P(\epsilon_{i}) = p\delta (\epsilon_{i}-1)+(1-p)\delta
(\epsilon_{i})$.   
As shown in Fig.~\ref{fig:flow-site}, there exists a random fixed
point, which is consistent with the universality scenario observed in
this $3D$ site-diluted  Ising model\cite{BFMMRG}. 
In the MC simulation\cite{BFMMRG}, the correction to the scaling becomes 
smaller around $p=0.80$. 
This can be explained by the finding that the random fixed point we found is
located on the position near the bare coupling with $p=0.80$. 
An interesting point to note is that the position of the fixed point is
close to that observed in the $3D$ bond-diluted Ising model, though
the corresponding bare parameters such as the concentration $p$ and the 
temperature differ between these two models, 
as seen in Fig.~\ref{fig:pd-dilute}. 
This finding implies that
the random fixed point is universal for a large class of the $3D$
unfrustrated random Ising models.

Next we consider a $3D \pm J$ Ising model, where the
interactions $J_{ij}$ are randomly distributed according to the 
bimodal distribution, $P(J_{ij}) =   p\delta (J_{ij}-J)+(1-p)\delta
(J_{ij}+J)$.  
The multi-spin coding technique can be easily implemented in this
model. For that purpose, we consider $32$ temperature points in the exchange MC
simulation. 
In this model, the ferromagnetically ordered state survives at
low temperatures up to a critical concentration, recently estimated at  
$p_{\rm c}=0.7673(3)$\cite{non-eq}, while below $p_{\rm c}$ a spin glass 
(SG) phase appears.
We estimate the domain-wall free energy with $L=8$ and $12$ for a wide
range of concentration $p$ including the critical concentration. 
Sample averages are taken over $240-6800$ samples depending on the size
and the concentration. 
\begin{figure}
\epsfxsize=\figwidth
\epsffile{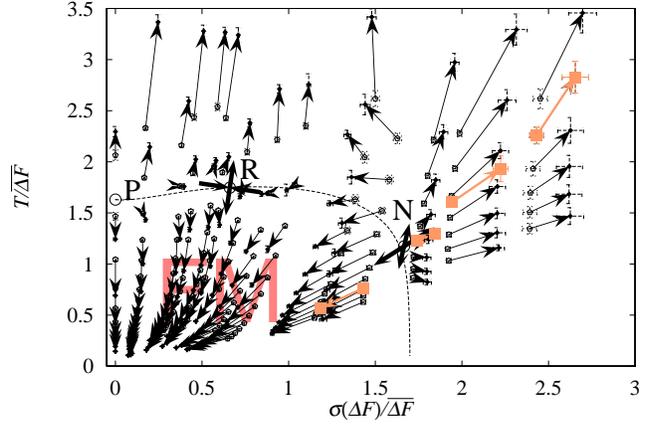}
\vup
\caption{
RG flow diagram for the  $3D \pm J$ Ising SG model. 
The pure critical fixed point is denoted by $P$, the random fixed
 point by $R$ and the multicritical fixed point by $N$. 
The bold arrows represent the eigenvector of the RG transformation
 matrix $\hat{T}$ at $R$ and $N$. 
The broken line represents a critical surface as a guide for the eyes. 
The random fixed point $R$ at $(0.66(2),1.74(1))$ is characterized by the
 eigenvalues $y_1=1.52(2)$ and $y_2=-0.42(13)$. 
The exponents associated with the fixed point $N$ at $(1.66(6),1.17(2))$ are
 $y_1=1.2(2)$ and  $y_2=0.62(6)$. 
Data points on the Nishimori line are marked by gray symbols. 
}
\vup
\label{fig:flow-sg3d}
\end{figure}

We show in Fig.~\ref{fig:flow-sg3d} the RG flow diagram for the $3D \pm
J$ Ising SG model. The arrow connects results for $L=8$ and $L'=12$. 
Here, we again find the random fixed point $R$ as an attractor along the
ferromagnetic phase boundary. The position of $R$ is also close to that
observed in the diluted Ising models. 
This finding indicates that both belong to the same universality class as
the ferromagnetic phase transition. 
In other words, the renormalized coupling constants (or coarse-grained
spin configurations) are independent of the details of the microscopic
Hamiltonian and of whether it is frustrated or not. 
One of the inherent characteristics of the $\pm J$ model is
the existence of a highly symmetric line, which we call the Nishimori
line\cite{Nishimori}.  There were several theoretical works concerning 
the Nishimori line\cite{Nishimori,DH,DH2,HTE,HTE2}.  
It was suggested by the $\epsilon$-expansion method and a symmetry
argument\cite{DH,DH2} that  the multicritical point must be located on
the line. 
This was confirmed by MC simulations\cite{non-eq,ON} and
high-temperature series expansions\cite{HTE,HTE2}. As shown in
Fig.~\ref{fig:flow-sg3d}, we also observe a fixed point, denoted by $N$,
corresponding to the multicritical point, where both scaling axes have
a positive eigenvalue.  
One of the scaling axes governed by the larger eigenvalue at $N$ almost
coincides with the Nishimori line, which is in agreement with the
predictions by the $\epsilon$-expansion\cite{DH,DH2}.  

\begin{figure}
\epsfxsize=\figwidth
\epsffile{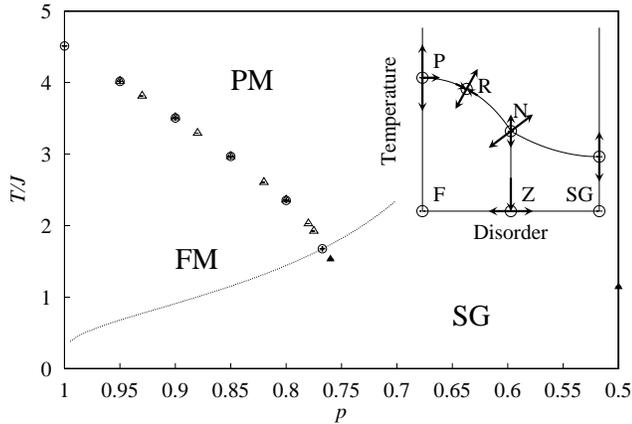}
\vup
\caption{
Phase diagram of the $3D \pm J$ Ising SG model.  Open triangles are
 transition temperature estimated 
in the present work.  
The SG transition temperatures marked by the solid triangles  at
 $p=0.50$ and $0.76$ are determined from  $\sigma(\Delta F)$.  
The open circles represent estimations by the nonequilibrium MC relaxation
 method{\protect\cite{non-eq,non-eq2}}.  
The dotted line represents the so-called Nishimori line. 
A schematic flow diagram for the $\pm J$ model is drawn in the inset. 
The fixed points $R$ and $N$, and the RG flow around them are supported
 in the present work.
}
\vup
\label{fig:phase}
\end{figure}

As seen in Fig.~\ref{fig:phase}, the estimated critical temperatures for
the concentrations simulated are  consistent  with those obtained by
the large-scale MC simulations up to size $L=101$\cite{non-eq,non-eq2}. 
Our analysis can also be performed for the SG transition at smaller
values of $p$. 
The SG transition temperature $T_{\rm SG}$ is determined from the
crossing of $\sigma (\Delta F)$. 
The value of $T_{\rm SG}$ at $p=0.50$ is in good agreement with 
the recent estimation\cite{KY}. 
Along the ferromagnetic phase boundary, it is natural to expect that
there exists another fixed point $Z$ at zero temperature, which
separates the ferromagnetic phase from the SG  one. 
Because the eigenvalues of the RG matrix at $N$ are positive for both
scaling axes, the thermal axis would be irrelevant in contrast to the
diluted Ising models, namely, an irrelevant flow into the
zero-temperature fixed point $Z$ originates at $N$, as shown in the
inset of Fig.~\ref{fig:phase}. 
The zero-temperature fixed point governs the critical behavior 
between the ferromagnetic and SG phase. 
In the present study, however, we cannot reach temperatures that are 
sufficiently low to extract the zero-temperature fixed point. 
A modern optimization technique such as a genetic algorithm 
would be useful in searching for the location of the fixed point.

Now we comment on the correction to the finite-size scaling. 
In order to obtain the critical exponent, one uses a range of the system
size frequently independent of the bare parameters in the model
Hamiltonian. 
When the system has a subleading scaling parameter, 
it causes systematic corrections to the scaling. 
However, it is not known how
such corrections affect the leading scaling {\it a priori}.  
The RG flow diagram gives  a good indication of necessity and
justification for corrections to the leading scaling term. 
Once the fixed-point structure of interest is clarified from the flow,
one of the best numerical approaches to obtain the critical exponents, within
restricted computer facilities, is to choose a model parameter close to
the fixed  Hamiltonian. 

In experiments, the $3D \pm J$ Ising model is approximately realized in
a mixed compound with strong anisotropy such as
Fe$_{x}$Mn$_{1-x}$TiO$_3$. 
Our findings suggest that the critical phenomena of the compound
Fe$_{x}$Mn$_{1-x}$TiO$_3$  near $x\sim 1$ belong to the same
universality class as the $3D$ random Ising models.  
We expect that various random Ising systems such as a mixed Ising spin
compound Fe$_{x}$Co$_{1-x}$F$_{2}$\cite{Nash}, though beyond the scope
of he present study, are also categorized into the same universality
class.  

In conclusion, 
we have studied the fixed-point structure of the $3D$ random-bond Ising
models using the numerical DWRG method. 
We have found in the $\pm J$ Ising model, a random fixed point besides the
pure and multicritical ones along the ferromagnetic phase boundary. 
Furthermore, the observed random fixed point is found to be very close
to  that of the $3D$ non-frustrated dilute Ising models, while bare
parameters in the model Hamiltonians differ entirely. 
This fact strongly suggests that there exists a universal fixed point
characterizing the $3D$ disordered ferromagnetic Ising model
irrespective of the type of disorder.  
The present work is, to our knowledge, the first of its kind performed
for determining the fixed-point structure of a $3D$ random spin system. 
We consider that the present numerical RG approach is quite useful
for understanding the universality class of random spin systems,
including spin glasses.

The author would like to thank H.~Takayama, K.~Nemoto and M.~Itakura for
helpful discussions. 
Numerical simulations have been performed mainly on Compaq and SGI
workstations at the Supercomputer Center, Institute of Solid State
Physics, University of Tokyo. 
Part of the simulations has been performed on Fujitsu VPP500 at the
Supercomputer Center, ISSP, University of Tokyo and Hitachi SR2201 at
the Supercomputer Center, University of Tokyo. 
The present work is supported by a Grant-in-Aid for the Encouragement of
Young  Scientists from the Ministry of Education (No. 11740220),
Science, Sports and Culture of  Japan.


\clearpage

\end{document}